\documentclass[sigconf,screen]{acmart}

\setcopyright{acmcopyright}
\acmPrice{15.00}
\acmDOI{10.1145/3533767.3534402}
\acmYear{2022}
\copyrightyear{2022}
\acmSubmissionID{issta22main-p430-p}
\acmISBN{978-1-4503-9379-9/22/07}
\acmConference[ISSTA '22]{Proceedings of the 31st ACM SIGSOFT International Symposium on Software Testing and Analysis}{July 18--22, 2022}{Virtual, South Korea}
\acmBooktitle{Proceedings of the 31st ACM SIGSOFT International Symposium on Software Testing and Analysis (ISSTA '22), July 18--22, 2022, Virtual, South Korea}

\usepackage{etoolbox,environ}

\usepackage{listings}
\usepackage{xcolor}
\usepackage{xspace}
\usepackage{enumitem}
\usepackage{hyperref}
\usepackage{tikz-uml}
\usepackage{microtype}
\usepackage{tabu}
\usepackage{multirow}
\usepackage{threeparttable}
\usepackage{pifont}
\usepackage[justification=centering]{caption}
\usepackage{setspace}
\usepackage{supertabular,booktabs}
\usepackage{longtable}
\usepackage[linesnumbered,ruled,lined]{algorithm2e}
\usepackage{makecell}
\usepackage{amsmath}
\usepackage{graphicx}
\usepackage{tcolorbox}
\usepackage{caption}
\usepackage{subfigure}
\usepackage{amsthm}

\newcommand{\tool}{{iFixDataloss}\xspace}
\newcommand*{\rom}[1]{\expandafter\@slowromancap\romannumeral #1@}

\usepackage{tabularx,adjustbox,booktabs}
\usepackage{blindtext}

\newcommand\clearrow{\global\let\rowmac\relax}
\usepackage{tikz}

\definecolor{mycolor}{rgb}{0.122, 0.435, 0.698}
\definecolor{gray1}{gray}{0.3}
\newcommand{\result}[1]{%
\begin{tcolorbox}[colframe=gray1,boxrule=0.5pt,arc=4pt,
      left=6pt,right=6pt,top=6pt,bottom=6pt,boxsep=0pt,width=\columnwidth]%
      {#1}
\end{tcolorbox}%
}

\newcommand{\todo}[1]{}
\renewcommand{\todo}[1]{{\color{red} TODO: {#1}}}

\begin{document}

\title{Detecting and Fixing Data Loss Issues in Android Apps}
\author{Wunan Guo}
\authornotemark[2]
\email{wnguo17@fudan.edu.cn}
\affiliation{
    \institution{Fudan University}
    \country{China}
}

\author{Zhen Dong}
\email{zhendong@fudan.edu.cn}
\authornotemark[2]
\authornote{Corresponding author.}
\affiliation{
    \institution{Fudan University}
    \country{China}
}

\author{Liwei Shen}
\authornotemark[2]
\email{shenliwei@fudan.edu.cn}
\affiliation{
    \institution{Fudan University}
    \country{China}
}

\author{Wei Tian}
\email{20212010141@fudan.edu.cn}
\authornote{Affiliated with the School of Computer Science and Shanghai Key Laboratory of Data Science, Fudan University, China.}
\affiliation{
    \institution{Fudan University}
    \country{China}
}

\author{Ting Su}
\email{tsu@sei.ecnu.edu.cn}
\authornote{Affiliated with Shanghai Key Laboratory of Trustworthy Computing.}
\affiliation{
    \institution{East China Normal University}
    \country{China}
}

\author{Xin Peng}
\authornotemark[2]
\email{pengxin@fudan.edu.cn}
\affiliation{
    \institution{Fudan University}
    \country{China}
}


\begin{abstract}
Android apps are \emph{event-driven}, and their execution is often interrupted by external events. This interruption can cause data loss issues that annoy users. For instance, when the screen is rotated, the current app page will be destroyed and recreated. If the app state is improperly preserved, user data will be lost. In this work, we present an approach and tool \tool that automatically detects and fixes data loss issues in Android apps. To achieve this, we identify scenarios in which data loss issues may occur, develop strategies to reveal data loss issues, and design patch templates to fix them. Our experiments on 66 Android apps show \tool detected 374 data loss issues (284 of them were previously unknown)  and successfully generated patches for 188 of the 374 issues. Out of 20 submitted patches, 16 have been accepted by developers.  In comparison with state-of-the-art techniques, \tool performed significantly better in terms of the number of detected data loss issues and the quality of generated patches.        

\end{abstract}



\maketitle

\section{Introduction}
\label{sec:intro}

A good user experience is essential for the success and popularity of mobile apps. On the other hand, a poor user experience can cause an app to become unpopular even if it provides useful functionality. Data loss is one of the prominent frustrations users can experience with mobile apps. Imagine a user is filling out a form with many fields and has almost completed the form after tediously typing through a tiny virtual keyboard. She or he might mistakenly touch the \emph{Back} button or be forcibly switched away from the app, causing all the input data to be lost. This occurs when an app page is destroyed by the mobile operating system. For example, when the mobile operating system needs to conserve resources or the mobile operating system determines that the app page is no longer needed (e.g., exiting via pressing the \emph{Back} button) or the app has been stored in the background for a long time. In such cases, if the app does not properly save user data, data loss issues will occur. Recent studies show data loss issues are extremely pervasive. Amalfitano et al.~\cite{amalfitano} reported that 60 out of the studied 68 apps (88.2\%) had data loss issues.

Fixing data loss issues automatically is non-trivial. A straightforward way to fix data loss issues is to save the values of all the state variables in an app page before it is destroyed, and restore these values when revisiting the app page. The problem with this solution is data \emph{over-saving}.  An app page may comprise a large amount of data that is irrelevant to user input (e.g., widgets displaying text). Saving this irrelevant data could slow down the app because such data saving occurs in the UIthread. As suggested in the Android documentation~\footnote{\url{https://developer.android.com/training/articles/perf-anr\#Avoiding}}, “any method that runs on the UI thread should do as little work as possible" avoiding UI sluggishness.  Additionally, data loss issues can lead to app misbehavior. For instance, a \texttt{TextView} on the app page displays how many times the user visits the page. Assume the current value of the \texttt{TextView} is 5 and saved before the app page is destroyed. When the user revisits the app page, value 5 is retrieved and displayed, which is confusing since the user expects 6. These have been identified as crucial problems in fixing data loss issues in recent works~\cite{livedroid,krerror}.

 A recent work, LiveDroid~\cite{livedroid}, uses static analysis to reason about program variables and GUI properties that might be changed during user interactions and generate patches to save and restore their values at runtime to avoid data loss issues. Although a significant portion of variables is ruled out in static analysis, LiveDroid still suffers from the over-saving issue. As reported in the paper, it generates too many false positives, i.e., preserving variable values that should not be preserved. 
 This will reduce app performance and responsiveness due to the cost of saving unnecessary data.
 
In this paper, we present a technique, called \tool, which can automatically detect and fix data loss issues in Android apps whilst 
eliminating the over-saving issue. The key insight of our technique is that the scenarios in which data loss issues occur can be simulated by generating a particular event or composed events during testing. For instance, screen rotation, one of the most frequent data loss scenarios, can be triggered by executing an orientation change event at app runtime. Thus, we can detect data loss issues by testing each app page for these data loss scenarios, and checking if data is lost. To fix data loss issues, \tool only preserves data that exhibits loss issues during testing thereby avoiding the saving of unnecessary data. To minimize unnecessary data saving, we further identify data related to user input (e.g., values of \texttt{TextField} widgets) and only save this kind of data in the patches.  

Specifically, \tool combines static and dynamic analysis to detect data loss issues. It first builds an \emph{activity transition graph} by performing static analysis on an app under test. Guided by this graph, \tool mimics user actions to exercise the app with a guided exploration strategy that steers the exploration towards app pages that may be affected by data loss issues. For each app page, \tool tests it by executing a set of predefined events that generate data loss scenarios. According to the lifecycle of Android activity, we define a set of events or composed events that cover possible data loss scenarios. Thus, \tool can find  data loss issues more 
thoroughly. In contrast, existing techniques~\cite{dld,livedroid} only cover a portion of those scenarios. 

For patch generation, \tool not only fixes data loss issues that only impact the current app run but also issues that impact users across multiple runs. Most of the time, data loss issues only affect app usage in the current run, e.g., certain \texttt{TextView} values on the screen are lost after a change in screen orientation. This kind of data typically only needs to be preserved for the current run and is no longer needed after the run is terminated. For such data, \tool generates patches that save it in memory and retrieve it when needed. In certain cases, data that is lost needs to be preserved across runs, we call this persistent data e.g., a membership registration form that is fully filled by users but has not been submitted; even after the app is killed, it still needs to be saved. For this kind of data, \tool generates patches that save the data using a storage method that can be restored across app runs. To achieve this, we develop a strategy based on user data usage patterns to distinguish these two kinds of data. In other words, \tool fixes not only data loss issues that occur in a singular instance of an app run but also data loss issues across app runs. By contrast, the existing technique LiveDroid~\cite{livedroid} only fixes data loss issues that occur in a singular instance of an app run and cannot fix data loss issues across multiple runs.              

Our experiments show that \tool is effective in both detecting and fixing data loss issues in Android apps. We evaluated \tool on 66 Android apps and detected 374 data loss issues and out of them, 284 were previously unknown.  \tool outperforms the recent data loss detection techniques DLD~\cite{dld} and LiveDroid~\cite{livedroid} regarding the number of detected data loss issues. It also outperforms the data loss issue fixing tool LiveDroid~\cite{livedroid} in terms of the quality of generated patches, i.e., the number of over-saved variable values in patches. For 188 issues, \tool successfully generated patches to fix them. For unknown issues,  we submitted 20 patches generated by \tool to developers; 16 of these 20 have been accepted by developers with very positive feedback.

Overall, our contributions can be summarized as follows:
\begin{itemize}
    \item  We identify scenarios in which data loss may occur based on the Android life cycle and design  strategies to reveal data loss issues. Further, we develop patch templates for fixing data loss issues. We also implement our approach into a fully automated tool \tool for detecting and fixing data loss issues in Android apps. 
    \item We performed an extensive experiment in which we found 374 data loss issues in 66 Android apps and 284 issues that were previously unknown and successfully generated patches for 188 out of the 374 issues. Out of 20 submitted patches, 16 have been accepted by developers.
    \item To facilitate future research, we make our prototype tool \tool and the data set used in the experiment are available at \url{https://github.com/iFixDataLoss/iFixDataloss22}
\end{itemize}
\section{Motivating Example}
\label{sec:example}
\begin{figure}[t]
	\centering
	\subfigure[before back]{
	    \label{fig:example1}
		\begin{minipage}{0.44\linewidth}
		\centering
		\hspace{-0.34cm}\includegraphics[width=1.6in]{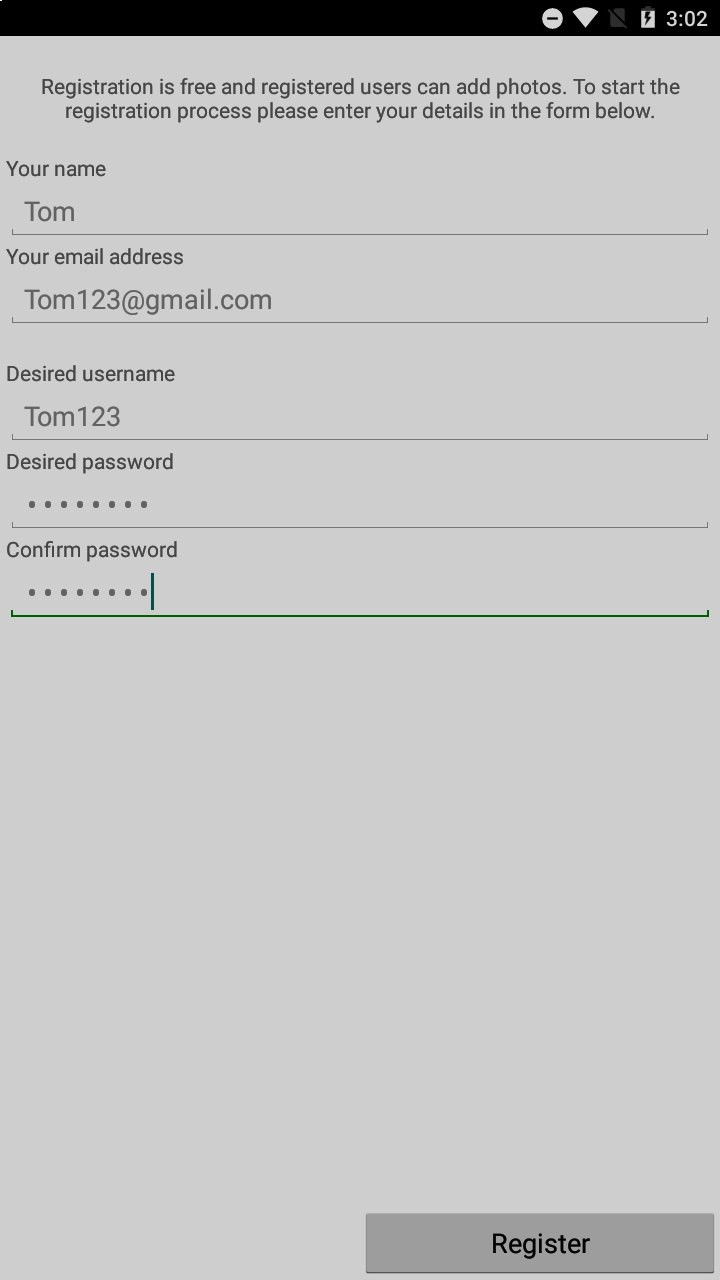}
		\end{minipage}
		
	}
\quad
	\subfigure[after back and return]{
	    \label{fig:example2}
		\begin{minipage}{0.44\linewidth}
		\centering
		\hspace{-0.34cm}\includegraphics[width=1.6in]{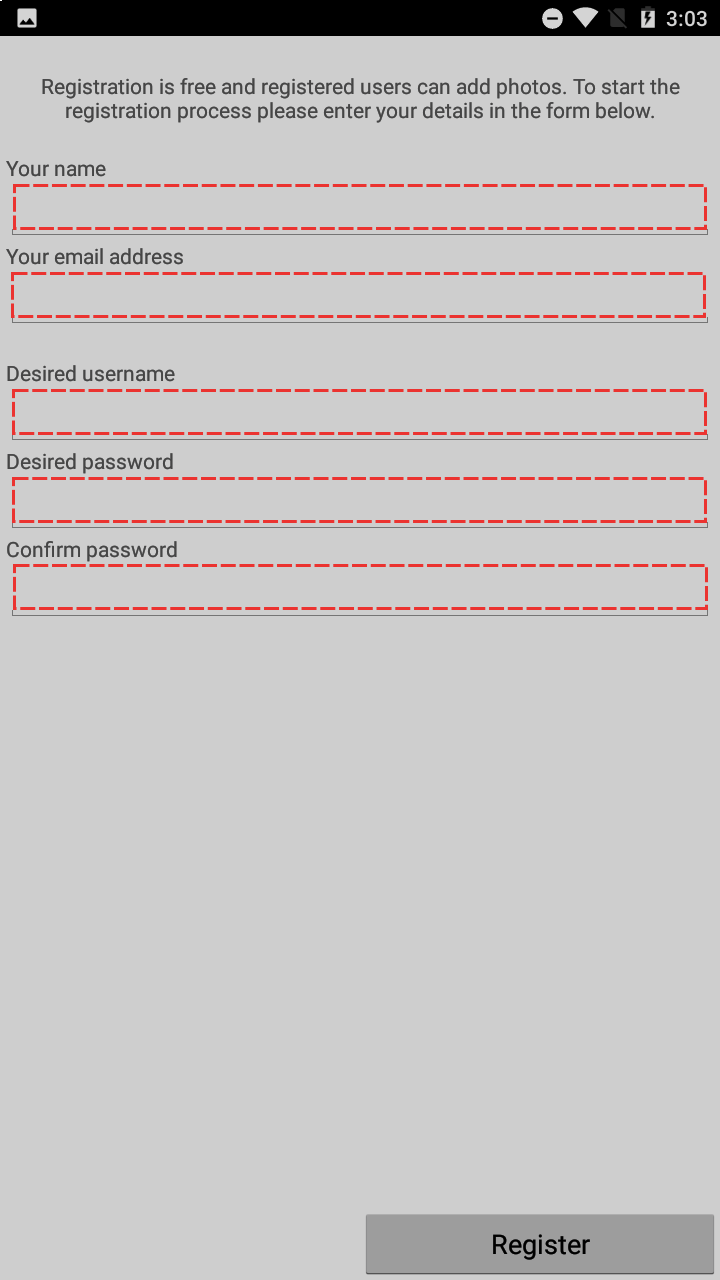}
		\end{minipage}
	}
	\caption{Screenshots of the CycleStreets activity described in the example.}
	\label{fig:motivation}
\end{figure}

In this section, we describe a data loss issue in a popular mobile app Cyclestreet (196 stars on Github and over 100K downloads on Google Play) as well as the 
inability of existing techniques in detecting and fixing this issue. Then we explain how it is detected and fixed by \tool. 

\paragraph{Data Loss.} Cyclestreets~\footnote{\url{https://www.cyclestreets.net/mobile/android/}} is a cycle journey planner mobile app that is widely used in the UK. \tool found a data loss issue in the app, which exists on its account registration page. 
Figure~\ref{fig:example1} shows the account registration process, which involves filling out the form using your personal information such as user name and email address. Subsequently, to finish registration, the user clicks the \emph{Register} button to submit the completed form. Upon investigation, there is a data loss issue in the page, which can be triggered in the following steps: (1) fill out the form without clicking the \emph{Register} button; (2) press the \emph{Back} button; (3) return to the registration page. As shown in Figure~\ref{fig:example2}, all the data filled in the previous step is lost after returning to the registration page. If users encounter this issue, it will frustrate users, since they have to refill this data which is tedious and time-consuming.       


\paragraph{Challenges.} There are difficulties in both detecting and fixing this data loss issue. Detecting such issues requires an oracle that checks whether user data is lost during testing. However, most existing mobile app testing tools \cite{sapienz,stoat,ape,timemachine,gesda,combodroid} can only detect crashes in apps and thus cannot be used to find and fix data loss issues. As mentioned earlier, a recent tool DLD~\cite{dld} is capable of detecting data loss issues. Unfortunately, DLD mainly detects data loss issues that occur when device orientation is changed and fails to detect the issue in the example. Automatically fixing data loss issues is non-trivial as well. 
The app page in the example comprises $17$ widgets and each widget contains more than $15$ properties, containing $311$ variable values in total. Restoring all of these values when entering the page can significantly slow down the app. We ran an experiment 10 times to compute the cost of saving and restoring all these variable values. 
On average, saving and restoring these 311 variables values takes 300ms, which is over 3 times the acceptable response time given in the Android documentation~\footnote{\url{https://developer.android.com/training/articles/perf-anr\#Reinforcing}}. Additionally, this time only takes into account saving and restoring. In a real-world app, there could be additional processing needed that would further increase the response time beyond the acceptable. To save less data, the state-of-the-art technique LiveDroid~\cite{livedroid} leverages static analysis to reason about variable values in GUIs that might be changed during user interaction and only saves them. 
In total, LiveDroid preserved 166 variable to fix this issue. Despite reducing the amount of data being saved, LiveDroid still exhibits an \emph{over-saving} issue. The majority of the 166 variable values are unnecessarily saved since they are initialization values and never changed during user interactions, e.g., \texttt{resource-id} and \texttt{content-desc}.

\begin{figure}[t]
	\centering
	\includegraphics[width=0.45\textwidth]{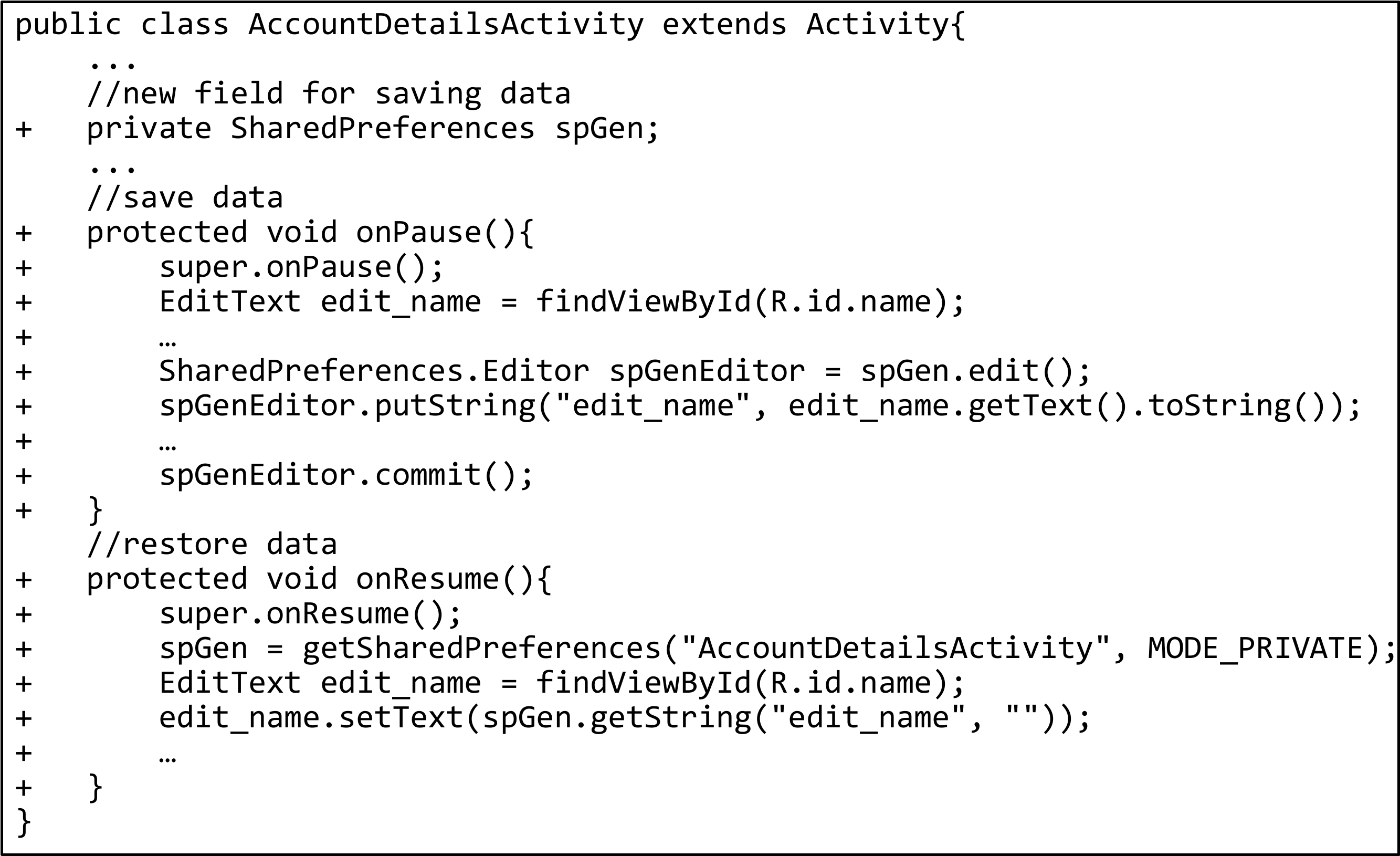}
	\caption{The patch generated by \tool for fixing the data loss issue in the example.}
	\label{fig:patch}
\end{figure}

\paragraph{Our approach.} The tool \tool, in which our approach is implemented, detected and fixed this issue via the following steps:
\begin{itemize}[leftmargin=*] 
    \item \emph{Data Loss Issue Detection.} \tool explores the app on an emulator and tests each discovered app page in data loss scenarios that we define based on the Android documentation. One of the scenarios is \texttt{Back-ReEntering}, i.e., exiting an app page via pressing the \texttt{Back} button and re-entering the page. In the example, \tool found that data filled in on the registration page was lost after the execution of the \texttt{Back-ReEntering} scenario. Thus, \tool reported the issue. During detection, \tool not only reports a data loss issue in an app page but also records variable values that are lost in a data loss scenario. Specifically, \tool modifies the values of variables on the page that may store user data such as \texttt{TextField} before generating a data loss scenario. If the value of a variable is changed in the data loss scenario, \tool deems that the variable has a data loss issue and its value needs to be saved. In the example, \tool found the values of these five \texttt{TextField} widgets (marked in red in Figure~\ref{fig:example2}) were changed in the \texttt{Back-ReEntering} scenario, and so the resource Ids of these widgets were recorded for patch generation in the next step. 
    
    \item \emph{Patch Generation.} \tool uses a template-based approach to generate patches. One template is designed to fix this kind of data loss issue that occurs in the example. With this template, \tool can generate a patch to fix this issue, in which only 5 variable values are preserved (shown in Figure~\ref{fig:patch}).
    
    
    \item \emph{Patch Evaluation.} \tool evaluates the generated patch by testing the patched APK. If data loss issues no longer exist in the corresponding app page and no crashes or freezes are found in the exploration of functionalities related to the app page, \tool reports the issue is fixed. In the example, \tool found neither data loss issues nor crashes or hangings in the evaluation and thus 
    determined that the patch was successful in fixing the issue. Furthermore, we created a pull request on the Github repo of the app Cyclestreet with the generated patch and the pull request was accepted. 
\end{itemize}
In summary, \tool successfully detected and fixed the data loss issue in the example. In the generated patch, only five widget values are saved and restored, and no unnecessary data is saved. The issue has been confirmed by the developers of the app Cyclestreets and the generated patch has been accepted as well.   



\section{Data Loss Revealing Strategy Design}
\label{sec:actdesign}
In this section, we first introduce the relevant basic concepts in the Android framework and analyze possible scenarios in which data loss issues could occur. At the end, we present the data loss revealing strategies we design for these scenarios.

\subsection{Basics}
\paragraph{Activity.} An activity is a fundamental component in Android that is used to implement an app page (also called screen) and contains the logic to handle the app page. Apart from activities, the Android framework provides \emph{fragments} that are smaller components for app page construction. An activity can have several fragments, each one containing both some graphical elements and the logic to handle them. Activities and fragments are the main components used in app page construction and are often affected by data loss issues. For the sake of simplicity, we refer only to activities in the following, but all our concepts apply to both activities and fragments. 

\paragraph{Activity Recreation.} This phenomenon frequently occurs during the execution of Android apps: an activity that a user is interacting with or put into the background can be destroyed due to system constraints (e.g., memory pressure or runtime configuration changes). When the user navigates back, the system recreates the activity using a set of saved data that describes the state of the activity when it was destroyed.       

\paragraph{Data Loss Issues.} Data loss is a state inconsistency issue that occurs in activity recreation, in which certain values of variables that describe the state are lost or assigned with initial values when recreating the activity that is destroyed earlier. This state inconsistency may cause inconvenience to users. Data loss often affects the GUI state, for instance, certain widgets may be missing their text values. In some cases, it may affect the internal state of the app. In this work, we mainly focus on data loss issues affecting the GUI state.  

\begin{figure}[t]
	\centering
	\includegraphics[width=0.5\textwidth]{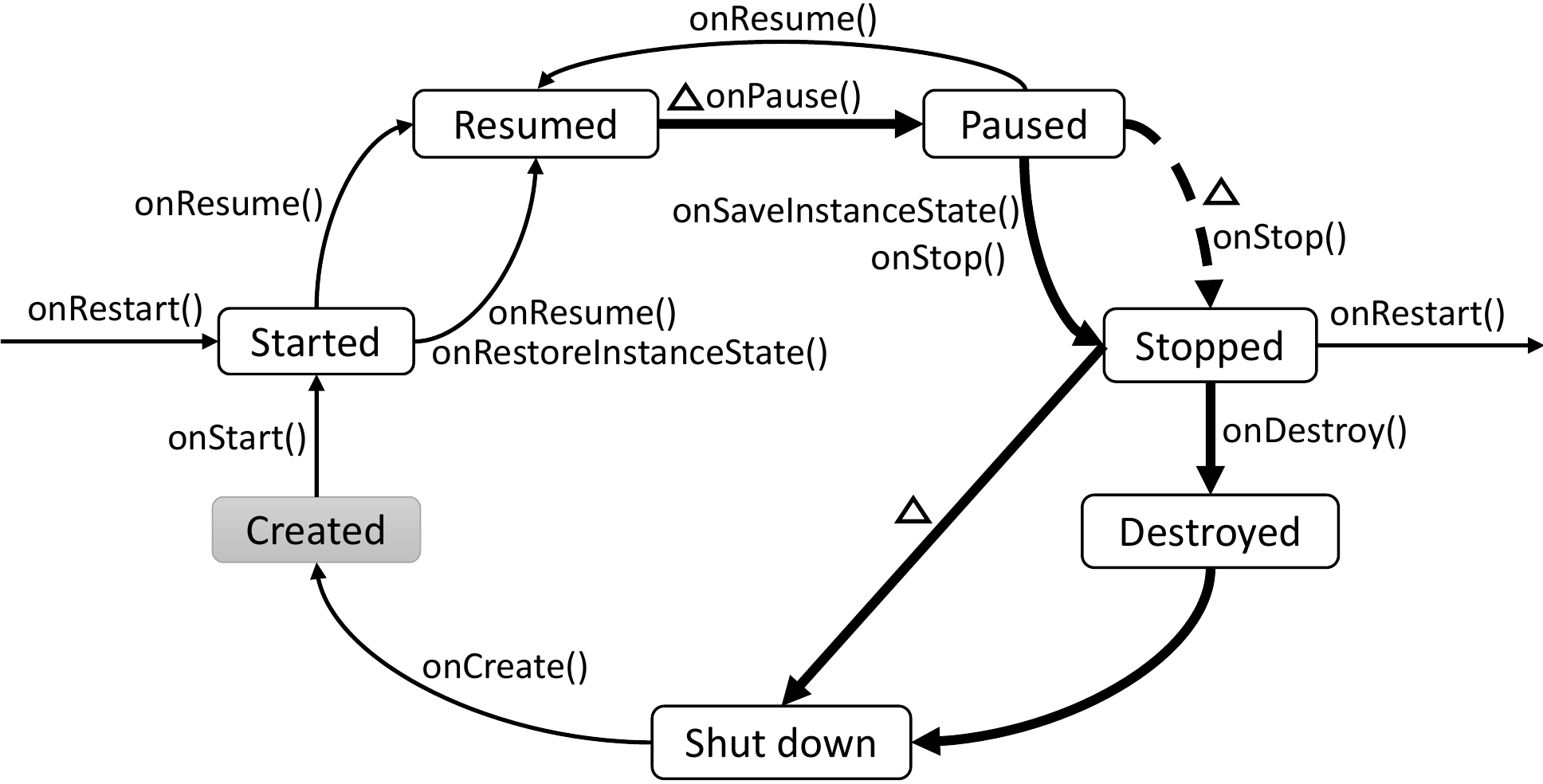}
	\caption{Activity lifecycle.}
	\label{fig:lifecycle}
\end{figure}

\subsection{Scenarios in Which Data Loss May Occur}

While activity recreation frequently occurs during app execution, the Android framework does not provide a fully automated mechanism for saving status data when an activity is destroyed and restored for activity recreation. App developers have to handle such status data during activity recreation. To support this, the Android framework provides callback methods (e.g., \texttt{onSaveInstanceState() and onRestoreInstanceState()}) that are invoked in the deconstruction and reconstruction of an activity, and developers can implement data saving and restoring functionality in these methods. However, an activity can be destroyed in multiple ways, for instance, being destroyed by pressing the \texttt{Back} button or being killed by the Android system for memory reclaiming. For each scenario, callback methods may be executed differently.  In the scenarios where developers don't implement or incorrectly implement the callback methods that handle status data, data loss issues will occur. 

As shown in the activity lifecycle in Figure~\ref{fig:lifecycle},  there are four state paths from state \texttt{Resumed} to \texttt{Shutdown}. That is, there are four theoretical activity lifecycle processes that the activity goes through when it is destroyed. For each process, a different sequence of callback methods is executed. However, in practice only three of them can occur in real-world scenarios. According to the official Android documentation\footnote{https://developer.android.com/guide/components/activities/activity-lifecycle}, the transition marked by the dashed line occurs only when the back button is pressed.  Subsequently, when the back button is pressed, the onDestroy() callback method is called. Therefore, the path marked by the triangle, cannot occur\footnote{Except in the rare case that this path is explicitly implemented by developers}. For the three scenarios that occur in practice, if developers do not properly save state data, data loss issues will happen. Thus, every activity in apps should be tested for these three scenarios:

\begin{itemize}[leftmargin=*] 

    \item \emph{Scenario 1 ($S1$):} \texttt{onPause()$\rightarrow$onSaveInstanceState()$\rightarrow$onSto\\p()}. This scenario often occurs when an activity is forcibly killed due to interrupt actions, for instance, being killed by the user swiping or killed by the Android system for memory reclaiming when it stays in the background for a long time.  
    
    \item \emph{Scenario 2 (S2):}  \texttt{onPause()$\rightarrow$onStop()$\rightarrow$onDestroy()}. This callback method execution order occurs when an activity is destroyed by the user pressing the \texttt{Back} button or the activity finishes itself. 

    \item \emph{Scenario 3 (S3):  \texttt{onPause()$\rightarrow$onSaveInstanceState()$\rightarrow$onSto\\p()$\rightarrow$OnDestroy()}}. This scenario occurs when an activity is destroyed and recreated due to runtime configuration changes e.g., screen rotation. For this case, status data in the activity needs to be completely saved in the deconstruction and restored in the recreation.
    
\end{itemize}
     
\subsection{Data Loss Issue Revealing Strategies}
In this section, we outline the three strategies we have designed to reveal data loss issues that occur in the three scenarios above.  A strategy can be expressed as a tuple $\langle E_d, E_r, O\rangle$, in which $E_d$ indicates an event sequence that destroys the current activity, $E_r$ indicates an event sequence that leads to re-entering of a state that was destroyed, and $O$ is a testing oracle that checks whether a data loss issue occurs during the activity recreation. 
The testing oracle $O$ is defined as
 \begin{definition}
     A data loss issue occurs when $V=\langle v_0, v_1 ..., v_n \rangle$ is different from $V^\prime=\langle v_0^\prime, v_1^\prime ..., v_n^\prime \rangle$, where $V$  represents values of variables within the GUI and $V^\prime$ represents their values after the activity recreation process.
\end{definition}
The idea behind these strategies is that given an app state, we execute events that trigger activity recreation and perform a state comparison to discover data loss issues. In the following, we introduce the three strategies:

\begin{itemize}[leftmargin=*] 

     \item $\mathcal{P}_{kill}:=\langle E_{kl}, E_{rp}, O_{kl}\rangle$. Strategy $\mathcal{P}_{kill}$ is designed to reveal data loss issues in scenario $S1$. $E_{kl}$ indicates an event sequence that simulates the user swiping action that kills the app. $E_{rp}$  represents the event sequence that restarts the app and re-enters the state when the app was killed. It can be recorded during state exploration (explained in Section~\ref{sec:approach}). $O_{kl}$ represents the testing oracle checking for data loss issues that occur in scenario S1. In $O_{kl}$, $V$ indicates variable values within the GUI that should be stored across app runs. This oracle is based on the suggestions shown in the box below, which is from an Android developer documentation~\footnote{\url{https://www.geeksforgeeks.org/shared-preferences-in-android-with-examples/}}. As suggested in the documentation, persistent data should be stored persistently to ensure a smooth user experience, even if the app is killed or restarted. Persistent data can be identified based on data usage patterns, e.g., stored in databases,  which will be further explained in Section~\ref{sec:approach}.
    \result{\emph{Persist data across user sessions, even if the app is killed and restarted, or the device is rebooted.}}

    \begin{table}[t]
	\caption{Editable Widgets.}
	\centering
	\begin{tabular}{|c|c|}
		\hline
		Type     & Widget Name                                                         \\ \hline
		Has-text & EditText, AutoCompleteTextView, Spinner                             \\ \hline
		No-text  & \multicolumn{1}{l|}{CheckBox, RadioButton, CheckedTextView, Switch} \\ \hline
	\end{tabular}
    \label{tab:editable_widgets}
\end{table}
    \item $\mathcal{P}_{back}:=\langle E_{bk}, E_{nx}, O_{bk}\rangle$. Strategy $\mathcal{P}_{back}$ is designed to reveal data loss issues in scenario $S2$. $E_{bk}$ indicates the event ``Back button'' and $E_{nx}$ represents an event sequence that re-enters the state that was just destroyed by pressing the Back button. The \texttt{Back button} is a system event that can be generated at any time and the event sequence $E_{nx}$ can be recorded during state exploration. $O_{bk}$ represents the testing oracle checking for data loss issues that occur in scenario S2. In $O_{bk}$, $V$ indicates property values of \emph{editable} widgets within the GUI. This oracle is derived from the suggestions shown in the box below, which is from the official Android developer documentation~\footnote{\url{https://developer.android.com/reference/android/app/Activity\#saving-persistent-state}}. As suggested in the document, the values of editable widgets should be preserved when the Back button is pressed. Editable widgets are listed in Table~\ref{tab:editable_widgets}.
    
    \result{\emph{The user pressing BACK from your activity does not mean "cancel" -- it means to leave the activity with its current contents saved away. Canceling edits in an activity must be provided through some other mechanism, such as an explicit "revert" or "undo" option.}}
    

    \item $\mathcal{P}_{rotate}:=\langle E_{rt},  $NOOP$,  O_{rt}\rangle$. Strategy $\mathcal{P}_{rotate}$ is designed to reveal data loss issues in scenario $S3$. $E_{rt}$ is a system event \texttt{screen-rotation}. When performing a screen-rotation event, the current state will be destroyed and recreated. Since activity recreation is automatically triggered in the screen rotation operation, the activity re-entering event is $NOOP$, i.e., a \texttt{do-nothing} event. $O_{rt}$ represents the testing oracle checking for data loss issues that occur in scenario S3. In $O_{rt}$, $V$ indicates values of all the widgets~\footnote{Here, only app-related properties are considered and screen-related properties such as widget bounds are excluded.} within the GUI. For this oracle, we consider property values of all widgets in the GUI because screen-rotation is a neutral event and its execution expects no change in the state.
 
\end{itemize}
Apart from the three oracles described above, \tool also detects other types of data loss issues with generic oracles such as crashes, and widgets disappearing.

\section{Approach}
\label{sec:approach}

\begin{figure}[h]
	\centering
	\includegraphics[width=0.48\textwidth]{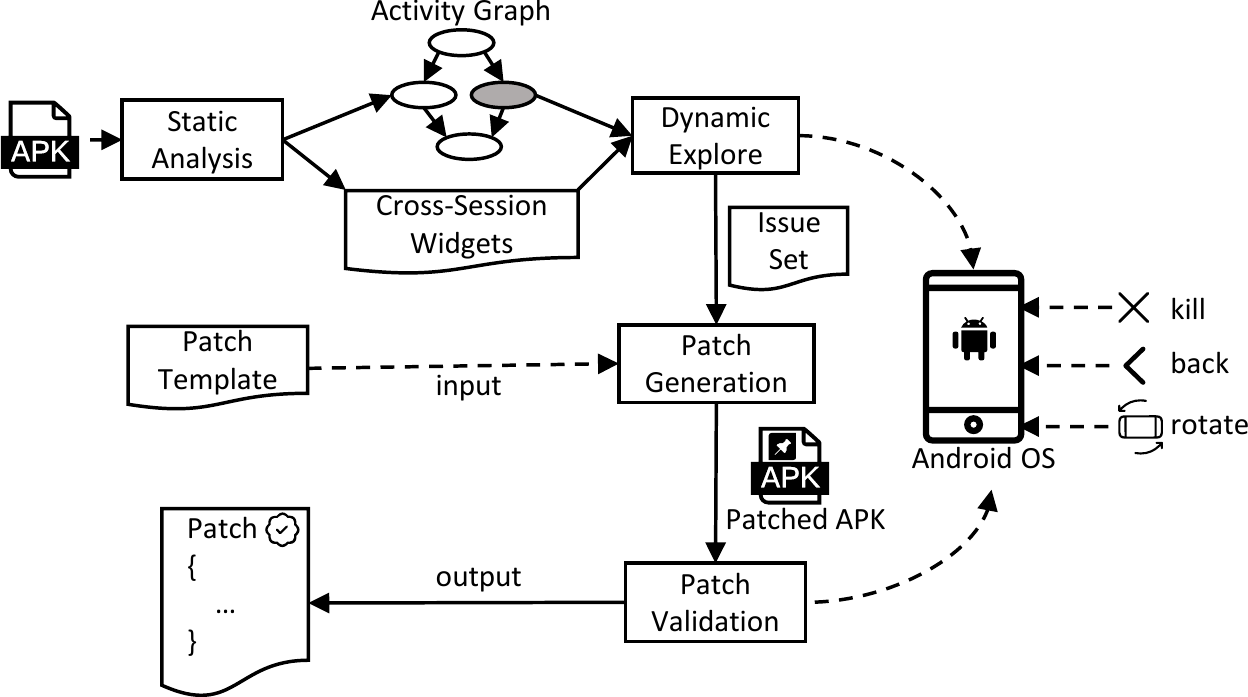}
	\caption{Workflow of our approach.}
	\label{fig:overview}
\end{figure}

The workflow of our approach implemented in \tool is shown in Figure \ref{fig:overview}. It comprises two steps: (1) data loss issue detection, (2) data loss issue fixing. Given an APK under test, \tool first uses static analysis and dynamic exploration to identify activities in the app that have data loss issues. For each activity that exhibits a data loss issue, \tool uses a template based approach to generate a patch to fix the data loss issue. Lastly, \tool runs the app and explores the patched activity to check if the data loss issue is correctly fixed.

\subsection{Data Loss Issue Detection.}
\begin{algorithm}[h]
  \small
  \caption{Data Loss Issue Detection \label{alg:detection}}

  \SetKwFunction{run}{Execute}

  \SetKwProg{proc}{Procedure}{}{}
  \SetKwFunction{build}{buildStaticActivityGraph}
  \SetKwFunction{infer}{inferWidgetStoreDataAcrossSessions}
  \SetKwFunction{explore}{guidedExplore}
  \SetKwFunction{launch}{launchAPP}
  \SetKwFunction{isExist}{isExist}
  \SetKwFunction{store}{put}
  \SetKwFunction{rotate}{testRotateStrategy}
  \SetKwFunction{restore}{restoreStatewithReplay}
  \SetKwFunction{back}{TestBackStrategy}
  \SetKwFunction{kill}{testKillStrategy}
  \SetKwFunction{getActivityId}{getActivityId}
   \SetKwFunction{change}{changeEidtableWidgetsValues}
  
  \KwIn{$APK$: App under test}
  \KwIn{$T_{ew}$: Set of editable widget types }
  \KwIn{$ART$: Android Runtime}
  \KwIn{$total\_time$: Time budget for testing}
  \vspace{5pt} 
  
  \proc{\run{$APK$, $T_{ew}$, $ART$, $total\_time$}} {
  \tcp{ACT\_id: Activity id}
  \tcp{$V$:Set of variables having data loss issues}
   $R\langle ACT\_id, V_{kl}, V_{bk}, V_{rt}, num\rangle \leftarrow \emptyset$ \;
   $Pool \leftarrow \emptyset$ \;
   \vspace{3pt} 
   \tcp{static analysis}
   $G \leftarrow \build()$ \;
   $W_{pr} \leftarrow \infer(T_{ew})$ \;
   
   \vspace{3pt} 
   \tcp{systematic testing}
   $ART \leftarrow \launch(APK, ART)$ \;
  \For{ $time < total\_time$}{
        $S\langle state\_id, Seq \rangle \leftarrow \explore(ART,G)$ \;
        \If{$isExist(S, Pool)$}{
            continue\;
        }
        $Pool \leftarrow \store(Pool, S)$ \;
        $ACT\_id \leftarrow \getActivityId(ART)$\;
        
        $\change()$\;
        $R \leftarrow \rotate(ART, ACT\_id)$ \; 
    
        $\restore(S,ART)$\;
        $\change()$\;
        $R \leftarrow \back(ART, T_{ew}, ACT\_id)$ \;
        
        $\restore(S,ART)$\;
        $\change()$\;
        $R \leftarrow \kill(ART, W_{pr}, ACT\_id)$ \;
            
        $\restore(S,ART)$\;
  
	 }
	 return $R\langle ACT\_id, V, num\rangle$ 
}
\end{algorithm}
Algorithm~\ref{alg:detection} outlines \tool's data loss issue detection procedure, which consists of two phases:\emph{static analysis} and \emph{systematic testing}. For the static analysis phase (Line 4-5), \tool extracts a static activity graph of the app under test that is used to guide exploration to quickly reach more activities. Additionally, \tool identifies variables within the GUI that should be stored persistently. 
In the systematic testing phase (Line 6-23), \tool runs the APK on an emulator and systematically explores the app. For each newly discovered state, \tool uses pre-defined strategies to create data loss scenarios and test the state in these scenarios to find data loss issues.



\subsubsection{Static Analysis}
\tool 's component for static analysis is built on FlowDroid~\cite{flowdroid}. FlowDroid is a widely used program analysis tool that can be used to perform data flow analysis for Android apps. Using FlowDroid, \tool is able to construct a static activity transition graph and identify persistent data within an app.

\paragraph{Static Activity Transition Graph.} The static activity transition graph in \tool is defined as $G=(A, E, \Sigma)$, where node $a \in A$ indicates an activity, edge $e \in E$ is an edge between nodes representing an activity transition, and label $\sigma \in \Sigma$ is a label on an edge representing a widget  with which associated events might trigger the activity transition. Here, we do not show our static activity graph construction algorithm in the paper due to limited space.   

\paragraph{Persistent Data Identification.} In practice, persistent data is typically saved on the internal storage via three storage solutions: \texttt{SharedPreferences}, SQLite databases, and the local file systems. \texttt{SharedPreferences} is a lightweight mechanism built into Android for saving and restoring data.  Adopting this solution, \tool identifies persistent data as follows: It analyzes the app code as well as the XML files, which describe GUI, to track the data flow of variables within the GUI. \tool reports variables whose values flow into invocations of APIs that are used to save persistent data. The APIs of saving persistent data are shown in Table \ref{tab:APIs}. \tool considers data stored in variables that match these criteria as persistent data. This practice is adopted in~\cite{krerror}  as well. 
\begin{table}[t]
	\caption{APIs for saving data across app runs.}
	\centering
\begin{tabular}{|l|l|}
\hline
\textbf{Class}                            & \textbf{Method} \\ \hline
Android.content.SharedPreferences         & put*            \\
Android.content.SharedPreferences\$Editor & put*            \\
Android.database.sqlite.SQLiteDatabase    & insert*         \\
Android.database.sqlite.SQLiteDatabase    & replace*        \\
Android.database.sqlite.SQLiteDatabase    & update*         \\
*OutputStream, *Writer                    & write*          \\ 
*                                         & save*           \\ \hline
\end{tabular}
    \label{tab:APIs}
\end{table}
\subsubsection{Systematic Testing}

\tool runs the app on an emulator and performs a systematic exploration, i.e., retrieving elements of GUI  and exercising them systematically. To test each state in data loss issue inducing scenarios, \tool implements the following modules:
\begin{itemize}[leftmargin=*]
    \item \emph{Transition Guided.} To quickly discover more states, \tool prioritizes exercising elements in GUI which are more likely to trigger activity transitions. To achieve this, \tool queries the static activity transition graph generated in the last step and localizes the elements in the current activity that might trigger transitions.
    
    \item \emph{Inputs Recording.} To test each state in all three data loss scenarios we designed, 
    \tool also records the sequence of events that lead to a specific state. This allows \tool to restore the app state by repeating the recorded event sequence.
    
    \item \emph{State Identification.}  To avoid repeatedly testing a state, \tool uniquely identifies a state by computing a hash over its widget hierarchy tree in which text-box values are removed (mitigating state explosion problem). This state abstraction is widely used in Android testing works~\cite{stoat,timemachine,ape}. As shown in Line 9-11, \tool skips the states that have been tested for data loss scenarios and only tests newly discovered states. 
\end{itemize}

As shown in Line 14-22 in Algorithm~\ref{alg:detection}, \tool tests each state for data loss issues in the three scenarios and records the variables that display data loss issues. The variables reported in scenario $\mathcal{P}_{rotate}$, $\mathcal{P}_{back}$ and $\mathcal{P}_{kill}$ are stored in set $V$, respectively. 
$num$ stores the total amount of times that other data loss issues, such as crashes, arise. In scenario $\mathcal{P}_{back}$, \tool requires identifying editable widgets, which is done by querying $T_{ew}$ storing editable widget types. In scenario $\mathcal{P}_{kill}$, \tool requires variables that have been identified as persistent data, which are stored in $W_{pr}$. In the end, \tool reports $R\langle ACT\_id, V, num\rangle$.

\subsection{Data Loss Issue Fixing}

In this section, we present templates used in patch generation and discuss how patches are evaluated in \tool. Note that, \tool focuses on fixing data loss issues where variable values are lost and leave out issues where a crash or hanging occurs in this work.     

\paragraph{Patch Templates.}  We fix data loss issues using templates derived from the official Android documentation. The recommended way to fix data loss issues is to save and restore data in \emph{proper} lifecycle methods with a \emph{proper} data saving mechanism. 
Following the suggestion in the documentation, we classify variables that have data loss issues into three categories and design patch templates accordingly:

\begin{itemize}[leftmargin=*]
     \begin{figure}[t]
		\centering
		\includegraphics{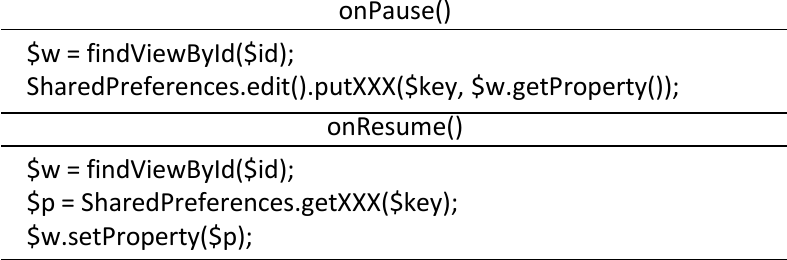}
		\caption{The patch template for preserving data across app runs.}
		\label{fig:template-cross-session}
	\end{figure}
    \item \emph{Storing values of editable widgets that need to be stored across app runs.} This kind of data is directly input by users and needs to be persistently saved. For example, edits that are made when composing an email. The Android documentation suggests saving this kind of data in \texttt{onPause()} method and restoring it in \texttt{onResume()} method to ensure nothing is lost in case the current activity is killed~\footnote{This callback is mostly used for saving any persistent state the activity is editing, to present a ``edit in place'' model to the user and making sure nothing is lost if there are not enough resources to start the new activity without first killing this one.}. To keep the methods \texttt{onPause()} and \texttt{onResume()} fast, we adopt the \texttt{SharedPreferences} framework to save and restore data since it is relatively lightweight compared to databases and file systems. The template for saving and restoring this category of values is shown in Figure~\ref{fig:template-cross-session}.
   
   	\begin{figure}[h]
		\centering
		\includegraphics{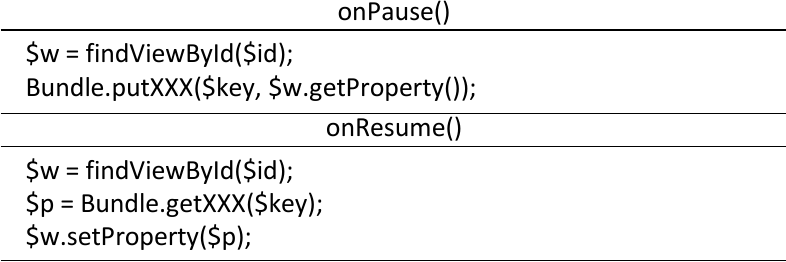}
		\caption{The patch template for preserving data in a single app run.}
		\label{fig:template-single-session}
	\end{figure} 
     \item \emph{Storing values of editable widgets that need to be stored for a single app run.} Consider the scenario where a user is using a Calculator app. If the user typed "33*23+3" in the \texttt{TextField}, this input should be saved for the duration of the computation session. However, this data should be cleared for the next run. As suggested in the documentation, we save and restore this kind of data in \texttt{onPause()} and \texttt{onResume()} methods, respectively. Since it is data for a single session, we save it in the \texttt{Bundle} object that is used for passing data between Android activities. The template for saving and restoring this category of values is shown in Figure~\ref{fig:template-single-session}.
    
    \begin{figure}[h]
		\centering
		\includegraphics{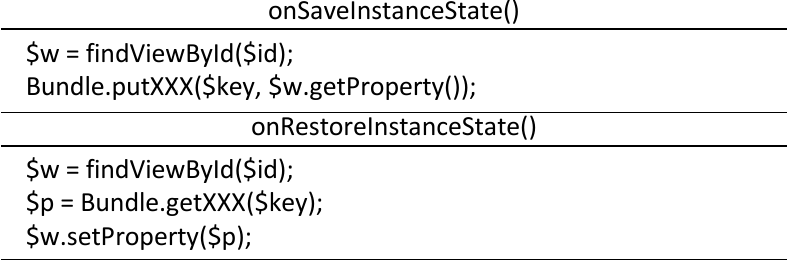}
		\caption{The patch template for preserving values of non-editable widgets.}
		\label{fig:template-non-editable}
	\end{figure}
     \item \emph{Storing values of non-editable widgets.} The data loss issue of non-editable widgets often occur during changes in runtime configuration, e.g., screen rotation. The documentation suggests saving and restoring them using \texttt{onSaveInstanceState()} and \texttt{onRestoreInstanceState()} methods. The template of them is shown in Figure~\ref{fig:template-non-editable}.
\end{itemize}

\begin{figure}[t]
	\centering
	\includegraphics[width=0.4\textwidth]{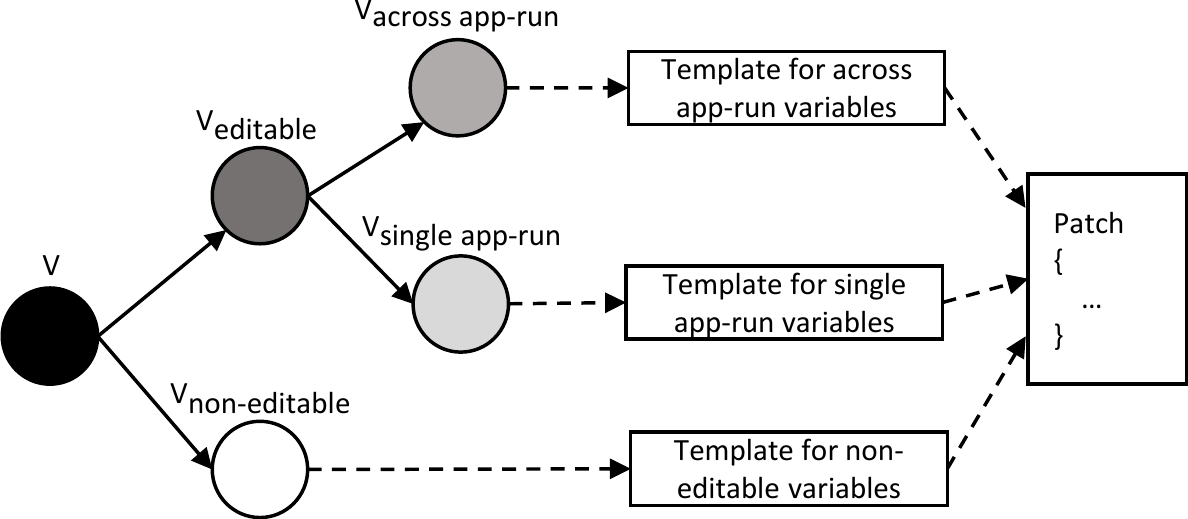}
	\caption{Workflow of patch generation.}
	\label{fig:patchgeneration}
\end{figure}

\paragraph{Patch Generation.} Figure~\ref{fig:patchgeneration} shows the workflow of the patch generation. Given a set of variables $V$ whose values need to be preserved,  \tool first divides them into two categories. The first category is the variables storing values of editable widgets and the second category is the variables storing values of non-editable widgets. The variables storing values of editable widgets are further divided into two categories: (1) variables whose values are used across app runs; (2) variables whose values are used in a single app run. In the end, these variables are divided into three categories as shown in Figure~\ref{fig:patchgeneration}. For each category of variables, \tool uses the corresponding template designed above to generate code for preserving their values and then it assembles code for preserving the three categories of variable values to generate a patch. Specifically, \tool converts the patch code to a set of \emph{AST} nodes and adds them into the \emph{AST} tree of the target activity's source code. 


\paragraph{Patch Evaluation.} To evaluate a patch, \tool runs the app and tests the patched activity in all three data loss scenarios to check if any data loss issue is found and then systematically explores functionality related to the activity to check if any crashes or freezing issues occur. If no error is found, \tool outputs the patch.

\section{Implementation}
\label{sec:impl}
\tool is implemented as a fully automated data loss detection and fix framework, which reuses or extends a set of off-the-shelf tools: ApkTool \cite{Apktool}, FlowDroid \cite{flowdroid}, UI Automator \cite{UiAutomator}, Android Debug Bridge(ADB) \cite{adb}, and JaveParser \cite{Javaparser}. Apktool is used to decompile an apk to extract its XML files. FlowDroid is extended to build the activity graph of an app and perform data flow analysis. UI Automator is used to dump the GUI layout and perform app state exploration. ADB is used to simulate the three revealing actions(i.e., kill, back and rotate). During patch generation, we use JaveParser to parse the source code of apps and generate patches. 
\section{Evaluation}
\label{sec:evaluation}

In our experimental evaluation, we seek to answer the following research questions:

\begin{itemize}[leftmargin=*]
    \item \textbf{RQ1}: How effective is \tool in finding data loss issues in Android apps? 
    \item \textbf{RQ2}: How is the quality of patches generated by \tool?
    \item \textbf{RQ3}: How useful is \tool in fixing real-world data loss issues in Android apps?
\end{itemize}

\subsection{Experimental Setup}

\paragraph{Subject Apps.} We evaluated \tool on a data set containing 66 Android apps, which is constructed by merging 48 benchmark apps used in Data Loss Detector~\cite{benchmark} and 21 apps that are found to have data loss issues in LiveDroid~\cite{livedroid}(there are 7 duplicate apps in these two benchmarks) as well as 4 apps downloaded from the Google Play Store. These 4 apps were used for data loss issues investigation during the early study  of our project and also included the evaluation data set (marked by ``\#'' in Table~\ref{tab:result}). Due to limited space, we use asterisks to omit some suffix letters of app names.   

\paragraph{Data Loss Issue Reporting. }  The experiments detect two types of data loss issues: GUI variables with missing values (indicated with $VE$) and critical errors such as crashing, hanging, and dialogues disappearing (indicated with $CE$). For each experiment,  we report the number of GUI variables that are not preserved during activity destruction and the number of critical errors.  For $VE$, we further classify them into two categories:
\begin{itemize}[leftmargin=*]
    \item \emph{True Positives (TP):} variables whose  values should be preserved. 
    \item \emph{False Positives (FP):} variables whose  values should not be preserved.
\end{itemize}
For each variable in $VE$, we manually check if the variable has a data loss issue. Specifically, we  explore the app and test the activity in which the variable resides in the following procedure. We first modify the values of all editable widgets and then perform a screen rotation operation and check if the variable remains the same before and after the screen rotation. Then we perform the procedure for the pressing Back button and Killing scenarios. If the variable value remains the same for the three scenarios, we deem the variable values should not be preserved, i.e. the variable is a false positive. Otherwise, the variable is a true positive. 


\paragraph{Comparison Tool Selection.} We compare \tool with LiveDroid~\cite{livedroid} and Data Loss Detector~\cite{dld}. LiveDroid is the most recent technique that prevents data loss issues in Android apps by automatically patching Android apps. 
Data Loss Detector is the most recent technique that detects data loss issues in Android apps. Data Loss Detector focuses on data loss issues caused by screen rotation and detects them by performing the screen rotation operation during testing. 

\paragraph{Execution Environment.} Our experiments run on a 64-bit Windows 10 physical machine with 2.30GHz Intel(R) Core(TM) i7-10510U CPU and 16GB RAM, and uses an Android emulator to run GUI exploration in the data loss detection phase. 
\begin{table}[t]
	\caption{Results of iFixDataloss, LiveDroid and DLD}
	\centering
	\resizebox{!}{100mm}{
		\begin{tabular}{|c|cccc|ccc|c|}
			\hline
			\multirow{2}{*}{\textbf{App}} & \multicolumn{4}{c|}{\textbf{iFixDataloss}}                                                           & \multicolumn{3}{c|}{\textbf{LiveDroid}}                           & \textbf{DLD} \\ \cline{2-9} 
			& \multicolumn{1}{c|}{\textbf{T}}   & \multicolumn{1}{c|}{\textbf{VE(TP)}} & \multicolumn{1}{c|}{\textbf{VE(FP)}} & \textbf{CE}  & \multicolumn{1}{c|}{\textbf{T}}   & \multicolumn{1}{c|}{\textbf{TP}} & \textbf{FP}  & \textbf{T}   \\ \hline
			Hourglass            & \multicolumn{1}{c|}{2}   & \multicolumn{1}{c|}{2}      & \multicolumn{1}{c|}{0}      & 0   & \multicolumn{1}{c|}{18}  & \multicolumn{1}{c|}{2}  & 16  & 1   \\
			K9                   & \multicolumn{1}{c|}{14}  & \multicolumn{1}{c|}{13}     & \multicolumn{1}{c|}{0}      & 1   & \multicolumn{1}{c|}{13}  & \multicolumn{1}{c|}{9}  & 4   & 1   \\
			QRStream             & \multicolumn{1}{c|}{3}   & \multicolumn{1}{c|}{0}      & \multicolumn{1}{c|}{0}      & 3   & \multicolumn{1}{c|}{2}   & \multicolumn{1}{c|}{0}  & 2   & 6   \\
			RingDroid            & \multicolumn{1}{c|}{5}   & \multicolumn{1}{c|}{3}      & \multicolumn{1}{c|}{0}      & 2   & \multicolumn{1}{c|}{1}   & \multicolumn{1}{c|}{1}  & 0   & 5   \\
			Browser              & \multicolumn{1}{c|}{1}   & \multicolumn{1}{c|}{0}      & \multicolumn{1}{c|}{0}      & 1   & \multicolumn{1}{c|}{20}  & \multicolumn{1}{c|}{0}  & 20  & 4   \\
			DroidShows           & \multicolumn{1}{c|}{3}   & \multicolumn{1}{c|}{2}      & \multicolumn{1}{c|}{0}      & 1   & \multicolumn{1}{c|}{6}   & \multicolumn{1}{c|}{1}  & 5   & 11  \\
			Gitclub              & \multicolumn{1}{c|}{3}   & \multicolumn{1}{c|}{0}      & \multicolumn{1}{c|}{0}      & 3   & \multicolumn{1}{c|}{0}   & \multicolumn{1}{c|}{0}  & 0   & 3   \\
			Glucosio             & \multicolumn{1}{c|}{24}  & \multicolumn{1}{c|}{23}     & \multicolumn{1}{c|}{0}      & 1   & \multicolumn{1}{c|}{0}   & \multicolumn{1}{c|}{0}  & 0   & 0   \\
			Gnucash              & \multicolumn{1}{c|}{6}   & \multicolumn{1}{c|}{5}      & \multicolumn{1}{c|}{0}      & 1   & \multicolumn{1}{c|}{1}   & \multicolumn{1}{c|}{1}  & 0   & 0   \\
			PGPClipper           & \multicolumn{1}{c|}{0}   & \multicolumn{1}{c|}{0}      & \multicolumn{1}{c|}{0}      & 0   & \multicolumn{1}{c|}{0}   & \multicolumn{1}{c|}{0}  & 0   & 0   \\
			MyDiary              & \multicolumn{1}{c|}{3}   & \multicolumn{1}{c|}{3}      & \multicolumn{1}{c|}{0}      & 0   & \multicolumn{1}{c|}{2}   & \multicolumn{1}{c|}{2}  & 0   & 0   \\
			Diary                & \multicolumn{1}{c|}{4}   & \multicolumn{1}{c|}{1}      & \multicolumn{1}{c|}{0}      & 3   & \multicolumn{1}{c|}{6}   & \multicolumn{1}{c|}{1}  & 5   & 5   \\
			Tuner                & \multicolumn{1}{c|}{6}   & \multicolumn{1}{c|}{0}      & \multicolumn{1}{c|}{0}      & 6   & \multicolumn{1}{c|}{0}   & \multicolumn{1}{c|}{0}  & 0   & 0   \\
			Rumble               & \multicolumn{1}{c|}{1}   & \multicolumn{1}{c|}{0}      & \multicolumn{1}{c|}{0}      & 1   & \multicolumn{1}{c|}{2}   & \multicolumn{1}{c|}{0}  & 2   & 1   \\
			LeafPic              & \multicolumn{1}{c|}{1}   & \multicolumn{1}{c|}{1}      & \multicolumn{1}{c|}{0}      & 0   & \multicolumn{1}{c|}{2}   & \multicolumn{1}{c|}{0}  & 2   & 0   \\
			OInotepad            & \multicolumn{1}{c|}{2}   & \multicolumn{1}{c|}{0}      & \multicolumn{1}{c|}{0}      & 2   & \multicolumn{1}{c|}{0}   & \multicolumn{1}{c|}{0}  & 0   & 0   \\
			TapeMeasure          & \multicolumn{1}{c|}{1}   & \multicolumn{1}{c|}{1}      & \multicolumn{1}{c|}{0}      & 0   & \multicolumn{1}{c|}{0}   & \multicolumn{1}{c|}{0}  & 0   & 0   \\
			androidclient        & \multicolumn{1}{c|}{2}   & \multicolumn{1}{c|}{2}      & \multicolumn{1}{c|}{0}      & 0   & \multicolumn{1}{c|}{4}   & \multicolumn{1}{c|}{2}  & 2   & 0   \\
			Notes                & \multicolumn{1}{c|}{1}   & \multicolumn{1}{c|}{1}      & \multicolumn{1}{c|}{0}      & 0   & \multicolumn{1}{c|}{1}   & \multicolumn{1}{c|}{1}  & 0   & 1   \\
			Tickmate             & \multicolumn{1}{c|}{13}  & \multicolumn{1}{c|}{0}      & \multicolumn{1}{c|}{0}      & 13  & \multicolumn{1}{c|}{0}   & \multicolumn{1}{c|}{0}  & 0   & 0   \\
			Timesheet            & \multicolumn{1}{c|}{4}   & \multicolumn{1}{c|}{4}      & \multicolumn{1}{c|}{0}      & 0   & \multicolumn{1}{c|}{0}   & \multicolumn{1}{c|}{0}  & 0   & 0   \\
			\#Webmon               & \multicolumn{1}{c|}{6}   & \multicolumn{1}{c|}{4}      & \multicolumn{1}{c|}{0}      & 2   & \multicolumn{1}{c|}{0}   & \multicolumn{1}{c|}{0}  & 0   & 2   \\
			\#ProExpense           & \multicolumn{1}{c|}{5}   & \multicolumn{1}{c|}{3}      & \multicolumn{1}{c|}{0}      & 2   & \multicolumn{1}{c|}{0}   & \multicolumn{1}{c|}{0}  & 0   & 1   \\
			\#Remembeer            & \multicolumn{1}{c|}{4}   & \multicolumn{1}{c|}{4}      & \multicolumn{1}{c|}{0}      & 0   & \multicolumn{1}{c|}{2}   & \multicolumn{1}{c|}{1}  & 1   & 0   \\
			\#arXivMobile          & \multicolumn{1}{c|}{7}   & \multicolumn{1}{c|}{6}      & \multicolumn{1}{c|}{0}      & 1   & \multicolumn{1}{c|}{9}   & \multicolumn{1}{c|}{3}  & 6   & 1   \\
			AntennaPod           & \multicolumn{1}{c|}{11}  & \multicolumn{1}{c|}{1}      & \multicolumn{1}{c|}{0}      & 10  & \multicolumn{1}{c|}{-}   & \multicolumn{1}{c|}{-}  & -   & 3   \\
			BeeCount             & \multicolumn{1}{c|}{5}   & \multicolumn{1}{c|}{5}      & \multicolumn{1}{c|}{0}      & 0   & \multicolumn{1}{c|}{3}   & \multicolumn{1}{c|}{1}  & 2   & 1   \\
			BookCatalogue        & \multicolumn{1}{c|}{14}  & \multicolumn{1}{c|}{13}     & \multicolumn{1}{c|}{0}      & 1   & \multicolumn{1}{c|}{4}   & \multicolumn{1}{c|}{2}  & 2   & 0   \\
			Calendar*            & \multicolumn{1}{c|}{1}   & \multicolumn{1}{c|}{0}      & \multicolumn{1}{c|}{0}      & 1   & \multicolumn{1}{c|}{-}   & \multicolumn{1}{c|}{-}  & -   & 0   \\
			Conversations        & \multicolumn{1}{c|}{5}   & \multicolumn{1}{c|}{5}      & \multicolumn{1}{c|}{0}      & 0   & \multicolumn{1}{c|}{-}   & \multicolumn{1}{c|}{-}  & -   & 0   \\
			CycleStreets         & \multicolumn{1}{c|}{6}   & \multicolumn{1}{c|}{5}      & \multicolumn{1}{c|}{0}      & 1   & \multicolumn{1}{c|}{166} & \multicolumn{1}{c|}{5}  & 161 & 3   \\
			DNS66                & \multicolumn{1}{c|}{4}   & \multicolumn{1}{c|}{4}      & \multicolumn{1}{c|}{0}      & 0   & \multicolumn{1}{c|}{1}   & \multicolumn{1}{c|}{1}  & 0   & 1   \\
			Document*            & \multicolumn{1}{c|}{1}   & \multicolumn{1}{c|}{1}      & \multicolumn{1}{c|}{0}      & 0   & \multicolumn{1}{c|}{-}   & \multicolumn{1}{c|}{-}  & -   & 2   \\
			Easy*                & \multicolumn{1}{c|}{4}   & \multicolumn{1}{c|}{1}      & \multicolumn{1}{c|}{0}      & 3   & \multicolumn{1}{c|}{-}   & \multicolumn{1}{c|}{-}  & -   & 6   \\
			Etar*                & \multicolumn{1}{c|}{2}   & \multicolumn{1}{c|}{1}      & \multicolumn{1}{c|}{0}      & 1   & \multicolumn{1}{c|}{0}   & \multicolumn{1}{c|}{0}  & 0   & 11  \\
			Firefox*             & \multicolumn{1}{c|}{1}   & \multicolumn{1}{c|}{1}      & \multicolumn{1}{c|}{0}      & 0   & \multicolumn{1}{c|}{-}   & \multicolumn{1}{c|}{-}  & -   & 0   \\
			Flym                 & \multicolumn{1}{c|}{9}   & \multicolumn{1}{c|}{3}      & \multicolumn{1}{c|}{0}      & 6   & \multicolumn{1}{c|}{5}   & \multicolumn{1}{c|}{1}  & 4   & 7   \\
			Gadgetbridge         & \multicolumn{1}{c|}{3}   & \multicolumn{1}{c|}{2}      & \multicolumn{1}{c|}{0}      & 1   & \multicolumn{1}{c|}{2}   & \multicolumn{1}{c|}{0}  & 2   & 2   \\
			LoopHabit*           & \multicolumn{1}{c|}{1}   & \multicolumn{1}{c|}{0}      & \multicolumn{1}{c|}{0}      & 1   & \multicolumn{1}{c|}{-}   & \multicolumn{1}{c|}{-}  & -   & 0   \\
			MALP                 & \multicolumn{1}{c|}{4}   & \multicolumn{1}{c|}{4}      & \multicolumn{1}{c|}{0}      & 0   & \multicolumn{1}{c|}{1}   & \multicolumn{1}{c|}{0}  & 1   & 5   \\
			MGit                 & \multicolumn{1}{c|}{4}   & \multicolumn{1}{c|}{0}      & \multicolumn{1}{c|}{0}      & 4   & \multicolumn{1}{c|}{3}   & \multicolumn{1}{c|}{0}  & 3   & 0   \\
			MTG*                 & \multicolumn{1}{c|}{40}  & \multicolumn{1}{c|}{29}     & \multicolumn{1}{c|}{0}      & 11  & \multicolumn{1}{c|}{2}   & \multicolumn{1}{c|}{0}  & 2   & 18  \\
			OmniNotes            & \multicolumn{1}{c|}{16}  & \multicolumn{1}{c|}{10}     & \multicolumn{1}{c|}{0}      & 6   & \multicolumn{1}{c|}{1}   & \multicolumn{1}{c|}{0}  & 1   & 0   \\
			OpenTasks            & \multicolumn{1}{c|}{11}  & \multicolumn{1}{c|}{11}     & \multicolumn{1}{c|}{0}      & 0   & \multicolumn{1}{c|}{1}   & \multicolumn{1}{c|}{1}  & 0   & 3   \\
			OpenVPN*             & \multicolumn{1}{c|}{1}   & \multicolumn{1}{c|}{0}      & \multicolumn{1}{c|}{0}      & 1   & \multicolumn{1}{c|}{0}   & \multicolumn{1}{c|}{0}  & 0   & 2   \\
			PassAndroid          & \multicolumn{1}{c|}{2}   & \multicolumn{1}{c|}{1}      & \multicolumn{1}{c|}{0}      & 1   & \multicolumn{1}{c|}{-}   & \multicolumn{1}{c|}{-}  & -   & 2   \\
			PeriodicTable        & \multicolumn{1}{c|}{1}   & \multicolumn{1}{c|}{0}      & \multicolumn{1}{c|}{0}      & 1   & \multicolumn{1}{c|}{0}   & \multicolumn{1}{c|}{0}  & 0   & 0   \\
			PortKnocker          & \multicolumn{1}{c|}{3}   & \multicolumn{1}{c|}{2}      & \multicolumn{1}{c|}{0}      & 1   & \multicolumn{1}{c|}{-}   & \multicolumn{1}{c|}{-}  & -   & 0   \\
			PrayerTimes          & \multicolumn{1}{c|}{3}   & \multicolumn{1}{c|}{0}      & \multicolumn{1}{c|}{0}      & 3   & \multicolumn{1}{c|}{-}   & \multicolumn{1}{c|}{-}  & -   & 0   \\
			QuasselDroid         & \multicolumn{1}{c|}{4}   & \multicolumn{1}{c|}{4}      & \multicolumn{1}{c|}{0}      & 0   & \multicolumn{1}{c|}{2}   & \multicolumn{1}{c|}{2}  & 0   & 0   \\
			QuickLyric           & \multicolumn{1}{c|}{5}   & \multicolumn{1}{c|}{1}      & \multicolumn{1}{c|}{0}      & 4   & \multicolumn{1}{c|}{2}   & \multicolumn{1}{c|}{0}  & 2   & 2   \\
			SimpleDraw           & \multicolumn{1}{c|}{0}   & \multicolumn{1}{c|}{0}      & \multicolumn{1}{c|}{0}      & 0   & \multicolumn{1}{c|}{0}   & \multicolumn{1}{c|}{0}  & 0   & 4   \\
			SimpleFile*          & \multicolumn{1}{c|}{2}   & \multicolumn{1}{c|}{0}      & \multicolumn{1}{c|}{0}      & 2   & \multicolumn{1}{c|}{-}   & \multicolumn{1}{c|}{-}  & -   & 4   \\
			SimpleGallery        & \multicolumn{1}{c|}{1}   & \multicolumn{1}{c|}{0}      & \multicolumn{1}{c|}{0}      & 1   & \multicolumn{1}{c|}{0}   & \multicolumn{1}{c|}{0}  & 0   & 0   \\
			SimpleSolitaire      & \multicolumn{1}{c|}{0}   & \multicolumn{1}{c|}{0}      & \multicolumn{1}{c|}{0}      & 0   & \multicolumn{1}{c|}{3}   & \multicolumn{1}{c|}{0}  & 3   & 4   \\
			Simpletask           & \multicolumn{1}{c|}{7}   & \multicolumn{1}{c|}{0}      & \multicolumn{1}{c|}{0}      & 7   & \multicolumn{1}{c|}{-}   & \multicolumn{1}{c|}{-}  & -   & 5   \\
			SMSBackup*           & \multicolumn{1}{c|}{0}   & \multicolumn{1}{c|}{0}      & \multicolumn{1}{c|}{0}      & 0   & \multicolumn{1}{c|}{0}   & \multicolumn{1}{c|}{0}  & 0   & 4   \\
			Syncthing            & \multicolumn{1}{c|}{13}  & \multicolumn{1}{c|}{0}      & \multicolumn{1}{c|}{0}      & 13  & \multicolumn{1}{c|}{0}   & \multicolumn{1}{c|}{0}  & 0   & 0   \\
			Taskbar              & \multicolumn{1}{c|}{14}  & \multicolumn{1}{c|}{0}      & \multicolumn{1}{c|}{0}      & 14  & \multicolumn{1}{c|}{0}   & \multicolumn{1}{c|}{0}  & 0   & 0   \\
			Tasks                & \multicolumn{1}{c|}{6}   & \multicolumn{1}{c|}{0}      & \multicolumn{1}{c|}{0}      & 6   & \multicolumn{1}{c|}{0}   & \multicolumn{1}{c|}{0}  & 0   & 11  \\
			Tusky                & \multicolumn{1}{c|}{2}   & \multicolumn{1}{c|}{1}      & \multicolumn{1}{c|}{0}      & 1   & \multicolumn{1}{c|}{1}   & \multicolumn{1}{c|}{1}  & 0   & 0   \\
			Twidere              & \multicolumn{1}{c|}{0}   & \multicolumn{1}{c|}{0}      & \multicolumn{1}{c|}{0}      & 0   & \multicolumn{1}{c|}{-}   & \multicolumn{1}{c|}{-}  & -   & 0   \\
			Vespucci             & \multicolumn{1}{c|}{28}  & \multicolumn{1}{c|}{0}      & \multicolumn{1}{c|}{0}      & 28  & \multicolumn{1}{c|}{-}   & \multicolumn{1}{c|}{-}  & -   & 0   \\
			VlilleChecker        & \multicolumn{1}{c|}{0}   & \multicolumn{1}{c|}{0}      & \multicolumn{1}{c|}{0}      & 0   & \multicolumn{1}{c|}{1}   & \multicolumn{1}{c|}{0}  & 1   & 4   \\
			WiFiAnalyzer         & \multicolumn{1}{c|}{14}  & \multicolumn{1}{c|}{0}      & \multicolumn{1}{c|}{0}      & 14  & \multicolumn{1}{c|}{1}   & \multicolumn{1}{c|}{0}  & 1   & 6   \\
			WorldClock*          & \multicolumn{1}{c|}{5}   & \multicolumn{1}{c|}{5}      & \multicolumn{1}{c|}{0}      & 0   & \multicolumn{1}{c|}{8}   & \multicolumn{1}{c|}{5}  & 3   & 0   \\ \hline
			Sum                  & \multicolumn{1}{c|}{374} & \multicolumn{1}{c|}{188}    & \multicolumn{1}{c|}{0}      & 186 & \multicolumn{1}{c|}{296} & \multicolumn{1}{c|}{43} & 253 & 152 \\ \hline
		\end{tabular}
	}
	\label{tab:result}
\end{table}
The emulator is configured with 2GB RAM and Android Nougat operating system(SDK 7.1, API level 25).  For each technique in the evaluation, we use the default parameter values given on its website. Regarding Data Loss Detector and \tool, we run each experiment for one hour.

\subsection{RQ1: Data Loss Issue Detection}

Table~\ref{tab:result} shows the results  of \tool, LiveDroid and Data Loss Detector 
after running through the 66 Android apps. Column ``T'' indicates the total number of detected data loss issues. Column ``TP'' and ``FP'' categorizes whether the reported issues are true positives or false positives. It is worth noting that, TP and FP for DLD are not shown in Table~\ref{tab:result} because in the comparison we focused on determining if a variable value detected by the tools is supposed to be preserved (described in Section-6.1). DLD only detects data loss issues and does not specify variable values that should be preserved and thus does not have a TP and FP column. Column ``CE'' indicates the number of critical errors detected by \tool. There are 14 apps in the data set that LiveDroid failed to run due to some compatibility issues, which are denoted by ``-'' in the table.

\begin{figure}[h]
	\centering
	\includegraphics[width=0.35\textwidth]{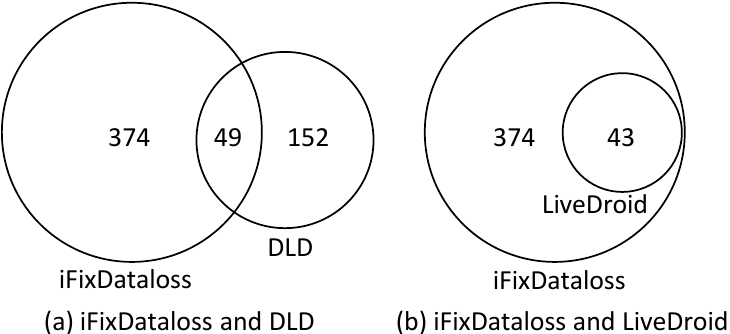}
	\caption{Data loss issues comparisons between \tool and the other two data loss issue detection tools.}
	\label{fig:intersection}
\end{figure}

\emph{Results.} \tool detected 374 data loss issues in the 66 Android apps. 188 out of these 374 issues are GUI variables with missing values and none were false positives. 
The remaining 186 issues are critical errors. Our investigation shows that out of the 374 data loss issues, 284 of them were previously unknown. In comparison with the state-of-the-art techniques, \tool detected the most data loss issues, followed by DLD (152) and LiveDroid (43). Regarding false positives, LiveDroid detected 296 GUI variables with missing values, but 253 are false positives. 
LiveDroid detects a significant amount of false positives because it uses static analysis to reason about variables whose values might change and suffers from the low precision problem in static analysis. We further perform a comparison  of detected data loss issues between \tool and state-of-the-art techniques. As shown in Figure \ref{fig:intersection}, \tool can detect all of the 43 issues that were detected by LiveDroid and 49 of 152 issues that were detected by Data Loss Detector. Two possible reasons that \tool could not detect all the issues are: (1) Data Loss Detector adopts screenshot-based oracles and reports a data loss issue whenever a difference is detected on the screenshot before and after screen rotation. Therefore it tends to report false positives, e.g.,  animations in an app page can cause differences on the screenshot even if there are no data loss issues in the app page; (2) \tool adopts a state exploration strategy different from Data Loss Detector and may miss certain states that were covered by Data Loss Detector. However, in total, \tool detected much more data loss issues than Data Loss Detector.

\result{\tool detected 374 data loss issues in the 66 Android apps. 284 out of 374 issues are previously undetected issues. \tool significantly outperforms the state-of-the-art techniques in terms of the number of detected data loss issues. }

\subsection{RQ2: Patch Quality}

\begin{table}[h]
	\caption{The distribution of patches generated by \tool and LiveDroid.}
	\centering
\begin{tabular}{|c|c|c|}
\hline
\textbf{Patch Type} & \textbf{iFixDataloss} & \textbf{LiveDroid} \\ \hline
Type 1              & 59                    & 7                  \\
Type 2              & 0                     & 10                 \\
Type 3              & 0                     & 6                  \\
Type 4              & 0                     & 21                 \\ \hline
\end{tabular}
    \label{tab:PatchType}
\end{table}
We evaluate the quality of a patch generated by \tool and LiveDroid based on two criteria. Firstly, we check if the patch successfully fixes the data loss issues without introducing new errors. For each patch, we run the patched app on an emulator to check if the data loss issue is fixed. Specifically, we manually test the patched activity in data loss scenarios and examine if the data loss issue still occurs. To ensure no new errors are introduced, we explore functionalities related to the patched activity and check if the app misbehaves, e.g. crashes or the GUI disappearing.  Secondly, we check if our patch generates any false positives i.e. variables in the GUI whose values should not be saved but saved because of the patch. 
Furthermore, we classify patches into four categories: 
\begin{itemize}[leftmargin=*]
    
    \item \emph{Type 1.} 
    Patch fixes all data loss issues without preserving false positives;
    \item  \emph{Type 2.} 
    Patch fixes all data loss issues but also preserves some false positives;
    \item  \emph{Type 3.} 
    Patch fixes some but not all data loss issues and also preserves some false positives;
    \item \emph{Type 4.} 
    Patch only preserves false positives.
    
\end{itemize}
Note that \tool is a fully automated tool and it can automatically evaluate patches as shown in Section~\ref{sec:approach}. In the experiment evaluation, we manually explore patched activities at runtime only for evaluating the quality of patches generated by \tool and LiveDroid.

 \paragraph{Results.} 59 patches were generated to address the 188 issues that involve variable values with missing detected by \tool. 
 Note that an activity may have multiple variables with data loss issues, in these cases, \tool generates one patch to preserve the values of all variables that exhibit data loss issues. As shown in Table~\ref{tab:PatchType}, all of the 59 patches generated by \tool fall into Type 1, i.e., all the 188 issues are fixed without preserving variable values that should not be preserved. In comparison, 296 issues that involve variables with missing values were detected by LiveDroid. LiveDroid generated 44 patches in total to address these issues. As shown in Table~\ref{tab:PatchType}, 7 of the 44 patches fall into Type 1, i.e., only 7 patches fix the data loss issues without saving and restoring variable values that should not be preserved. The remaining patches have the \emph{over saving} problem, i.e., saving and restoring variable values that should not be preserved. Out of the 296 variables values identified by LiveDroid, 253 are false positives, i.e., unnecessarily preserved. On average, 85\% (253/296) of the variable's values are values that should not be saved and exhibit the over-saving issue.
 
 \result{\tool generated 59 patches that successfully fixed 188 variable values miss issues without preserving false positives (variable values that should not be preserved). \tool outperformed the state-of-the-art technique LiveDroid in terms of the number of preserved false positives.}
\begin{table*}[h]
	\caption{Details of submitted patches.}
	\centering
	\resizebox{\textwidth}{!}{
		\begin{tabular}{|c|c|l|c|}
			\hline
			\textbf{App}         & \textbf{Activity} & \multicolumn{1}{c|}{\textbf{Issue Description}}                                                                    & \textbf{Status} \\ \hline
			A*                   & SearchWindow      & Journal article data will be lost when leaving without searching and returning to the article search page.         & Accepted        \\ \hline
			\multirow{2}{*}{Be*} & NewProject        & Project name will be lost when leaving without saving and returning to the create new project page.                & Accepted        \\ 
			& CountOptions      & Count data will be lost when leaving without saving and returning to the CountOptions page.                        & Accepted        \\ \hline
			\multirow{4}{*}{Bo*} & BookISBNSearch    & Book ISBN number will be lost when leaving without saving and returning to the ISBN search page.                        & Accepted        \\ 
			& EditAuthorList    & Author data will be lost when leaving without saving and returning to the add/edit author entry page.              & Accepted        \\ 
			& EditSeriesList    & Book series data will be lost when leaving without saving and returning to the add/edit book series page.          & Accepted        \\ 
			& BookEdit          & Book data will be lost when leaving without saving and returning to the add/edit book entry page.                  & Accepted        \\ \hline
			C*                   & AccountDetails    & Registration data will be lost when leaving without submitting and returning to the account creation page.         & Accepted        \\ \hline
			D*                   & Item              & Host data will be lost when leaving without saving and returning to the create host entry page.                    & No Response     \\ \hline
			Ga*                  & Debug             & Debug message will be lost when leaving without sending the message and returning to the debug page.               & Accepted        \\ \hline
			\multirow{6}{*}{Gl*} & AddGlucose        & Glucose data will be lost when leaving without saving and returning to the add glucose entry page.                 & Accepted        \\ 
			& AddA1C            & A1C data will be lost when leaving without saving and returning to the add A1c entry page.                         & Accepted        \\ 
			& AddWeight         & Weight data will be lost when leaving without saving and returning to the add weight entry page.                   & Accepted        \\ 
			& AddPressure       & Blood pressure data will be lost when leaving without saving and returning to the add blood pressure entry   page. & Accepted        \\ 
			& AddKetone         & Ketone data will be lost when leaving without saving and returning to the add ketone entry page.                   & Accepted        \\ 
			& AddCholesterol    & Cholesterol data will be lost when leaving without saving and returning to the add cholesterol entry page.         & Accepted        \\ \hline
			Gn*                  & AccountForm       & Expense account data will be lost when leaving without saving and returning to the add expense account page.                       & No Response     \\ \hline
			P*                   & Main              & Expense data will be lost when leaving without saving and returning to the save expense page.                      & No Response     \\ \hline
			T*                   & Calibration       & Calibration length will be lost after pressing back and returning to the calibration page.                         & No Response     \\ \hline
			W*                   & CreateEntry       & Website data will be lost when leaving without saving and returning to the create new entry page.                  & Accepted        \\ \hline
		\end{tabular}
	}
	\label{tab:PR}
\end{table*}
\subsection{RQ3: Usefulness }

To evaluate how useful \tool is in practice,  we selected the top 10 apps in the data set that have been updated most recently and submitted all the patches for these apps to developers (20 patches). In total, we created and submitted 20 pull requests for the 10 apps on Github. At the time of writing the paper, out of the 20 pull requests, 16 have been accepted with very positive comments:

\begin{itemize}[leftmargin=*]
    \item "Looks good - thanks again"
    \item "Super thank you for the contribution"
    \item "Thank you very much for this nice contribution. It looks really cool, overall."
    \item "Thank you! I have tested this and merged by rebasing into the master."
    \item ......
\end{itemize}
 We have not yet received a response for the remaining 4 pull requests. The details of the 20 patches are shown in Table~\ref{tab:PR}. Similar to Table~\ref{tab:result}, we used asterisks to omit some suffix letters of app names due to space limitations.
 


\result{Out of the 20 pull requests with patches generated by \tool, 16 have been accepted with positive comments.}

\subsection{Threats to Validity}
The main threats to external validity lie in the selection of the apps. 
\tool is evaluated on 66 Android apps. Our results may not be generalizable beyond the 66 apps to which we have applied \tool. To mitigate this threat, we chose apps from within the benchmarks of the two literature works, Data Loss Detector and Livedroid. Threats to internal validity are factored into our experimental methodology and they may affect our results. We manually explore apps to reproduce data loss issues reported in the experiments and may fail to explore certain functionalities, which might affect our results. 
We also performed some manual checks during evaluation, which is potentially error-prone. To minimize this threat, two people performed manual checks and compared the experimental results to check for discrepancies.
We also realise that the false positive rate of LiveDroid reported in our experiments is different from the rate reported in the LiveDroid paper. We checked with the authors of LiveDroid on this matter. The authors of LiveDroid explained that they also consider non-UI property values in the calculation of their false positive rate. Non-UI property values do not apply to our scenarios, therefore, resulting in a different false positive rate.

We used the default parameters when running LiveDroid and DLD. For LiveDroid, we tried running each app with different parameters and found there was no difference in results compared to the default parameters. For DLD, we tried running each app with different parameters except for the runtime length which we kept at 1 hour and found there was no difference in results compared to the default parameters. We had 2 people run these experiments to ensure that the results were consistent. Based on this, we concluded that running the experiments using default parameters did not affect the results.




\section{Related Work}
\label{sec:relatedwork}
\paragraph{Data Loss Issues Fixing.}  There have been a few pieces of research work that attempt to automatically fix data loss issues in Android apps. A recent work LiveDroid~\cite{livedroid} leverages static analysis to identify program variable values and GUI properties that may change during user interactions and preserve them across app life cycles to avoid data loss issues. Due to ``infeasible'' paths introduced in the static analysis, LiveDroid tends to report false positives in results. Furthermore, LiveDroid only works for data loss issues across app life cycles such as activity recreation and does not work for data loss issues across app runs. RuntimeDroid~\cite{runtimedroid} uses an online resource loading module to update GUI elements when certain configurations are changed at runtime, avoiding activity restarting. By contrast, \tool can fix data loss issues by preserving variable values whose loss issues have been witnessed during testing and restoring them in data loss scenarios. Thus, \tool generates no false positives. Apart from fixing data loss issues across app life cycles, \tool also can fix data loss issues across app runs.      

\paragraph{Automated Program Repair for Android Apps.} There are several existing works that focus on fixing other types of issues in Android apps. Droix~\cite{droix} employs a search-based approach to generate patches that fix crashes in Android apps. AppEvolve~\cite{appevolve} analyzes existing updates in other apps to generate patches that fix issues in Android apps caused by an API evolving. METER~\cite{meter} leverages computer vision techniques to fix broken GUI test scripts during app evolution. Sapfix~\cite{sapfix}, a deployed automated program repair tool in Facebook, seeks to generate patches for more types of bugs in mobile apps with templates that are created by human engineers based on previous bug fixes. Compared to those works, \tool focuses on fixing data loss issues in Android apps and can be used to complement those tools for fixing more types of issues in Android apps. 

\paragraph{Data Loss Issues Detection.} Similar to \tool, data loss issue detection tools DLD~\cite{dld} and ALARic~\cite{alaric} exercise app pages by executing a screen rotation action and detecting data loss issues by checking for differences in the GUI before and after rotation. Thor~\cite{thor} augments existing test suites with neutral sequences of operations to reveal more failures. The injected event sequences that create adverse conditions such as disconnecting the internet and turning off audio services, may reveal data loss issues. Quantum~\cite{quantum} tests Android apps by generating test cases that are injected with a series of operations that are more likely to reveal failures based on a study of previous bugs (e.g., zooming in and zooming out). SetDroid~\cite{setdroid} executes test cases under different system settings to find system setting related failures. Despite their ability to find data loss issues, those techniques are only able to find a portion of data loss issues. By contrast, we design operations that cover all kinds of scenarios in which data loss issues occur based on the Android lifecycle and use them to discover more types of data loss issues. More importantly, not only can \tool find data loss issues but it can also fix them.        

\paragraph{Automated Android App Testing.} Another rich branch of research work focuses on generating test inputs for Android apps. For instance, Sapienz~\cite{sapienz} uses evolutionary algorithms to generate test inputs that cover more code coverage. TimeMachine~\cite{timemachine} saves app states that have the potential to trigger new program behaviors and prioritizes exploring them to discover more app behaviors. Stoat~\cite{stoat}, APE~\cite{ape} and DroidBot~\cite{droidbot} leverages a built model to guide input generation. A3E systematically generates inputs following a depth-first strategy. SwiftHand~\cite{swifthand} uses machine learning to learn a model that is used to guide input generation. ACTEve~\cite{acteve} uses symbolic execution to generate inputs. While these techniques can effectively explore app behaviors, they are insufficient for detecting data loss issues due to a lack of oracles that check for data loss issues.

\section{Conclusion}
\label{sec:conclusion}
We introduced a practical technique that can automatically detect and fix data loss issues in Android apps and implemented it into a tool \tool.  Our extensive experiments (66 apps) show \tool is effective in detecting and fixing data loss issues in Android apps. In the evaluated 66 apps, \tool detected 374 data loss issues and 284 of them were previously unknown. \tool successfully generated 59 patches for 188 out of the 374 issues. Out of the 20 submitted patches, 16 have been accepted by developers. The experiments also show that iFixDataloss outperforms the state-of-the-art techniques in terms of the number of detected data loss issues and the quality of generated patches. To facilitate future research on data loss issues, we make our tool and data set publicly available at \cite{iFixDataloss-artifact}.
\section*{Acknowledgements}
This work is partially supported by National Natural Science Foundation of China (NSFC) under Grant No. 61972098. Ting Su is partially supported by NSFC Project No. 62072178.
\balance
\bibliographystyle{ACM-Reference-Format}
\bibliography{bibliography}

\end{document}